# CRYPTANALYSIS AND FURTHER IMPROVEMENT OF A BIOMETRIC-BASED REMOTE USER AUTHENTICATION SCHEME USING SMART CARDS

Ashok Kumar Das

Center for Security, Theory and Algorithmic Research
International Institute of Information Technology, Hyderabad 500 032, India
iitkgp.akdas@gmail.com

## ABSTRACT

*Recently, Li et al. proposed a secure biometric-based remote user authentication scheme using smart cards to withstand the security flaws of Li-Hwang's efficient biometric-based remote user authentication scheme using smart cards. Li et al.'s scheme is based on biometrics verification, smart card and one-way hash function, and it also uses the random nonce rather than a synchronized clock, and thus it is efficient in computational cost and more secure than Li-Hwang's scheme. Unfortunately, in this paper we show that Li et al.'s scheme still has some security weaknesses in their design. In order to withstand those weaknesses in their scheme, we further propose an improvement of their scheme so that the improved scheme always provides proper authentication and as a result, it establishes a session key between the user and the server at the end of successful user authentication.*

## 1. INTRODUCTION

In a client/server system scenario, usually a password-based authentication scheme with smart card is widely used in order to identify the validity of a remote user [1], [4], [10]. Traditional remote identity-based authentication schemes [2], [5] are based on passwords and thus the security of these schemes is based on the passwords only. But, simple passwords are always easy to break using simple dictionary attacks [6]. Thus, to overcome such problems, cryptographic secret keys and passwords are suggested to be used in the remote user authentication schemes. However, the long and random cryptographic keys are difficult to memorize and hence they must be stored somewhere, and it makes expensive to maintain the long cryptographic keys in practice. The drawback with remote user authentication schemes using cryptographic keys and passwords is that they are unable to provide non-repudiation because cryptographic keys and passwords can be forgotten, lost or even they may be shared with other people and as a result there is no way to know who is the actual user. Generally, a secure and efficient remote user authentication scheme should meet the following conditions [7]:

- Compatibility with a multi-server network architecture without repetitive registration.
- Low computational workload of the smart card.
- No need of password table or verification table.
- Resistance to different kinds of attacks.
- Allow the user to choose his/her identifier, password and update his/her password freely.

- Allow the users and servers to authenticate each other and then negotiate a session key to security communication.
- No requirement on time synchronization and delay-time limitation.

On the other hand, recent biometric-based remote user authentication schemes along with passwords have drawn considerable research attention [3], [6], [7], [8]. There are many advantages of using biometric keys (for example, fingerprints, faces, irises, hand geometry and palm-prints, etc.) as compared to traditional passwords, which are (as described in [6])

- Biometric keys cannot be lost or forgotten.
- Biometric keys are very difficult to copy or share.
- Biometric keys are extremely hard to forge or distribute.
- Biometric keys cannot be guessed easily.
- Someone's biometrics is not easy to break than others

Biometric-based remote user authentications are thus inherently more reliable and secure than usual traditional password-based remote user authentication schemes.

In 1981, Lamport first introduced the notion of remote authentication scheme [4] in which a remote server can authenticate a remote user based on identity and password over an insecure network. Lamport's scheme requires to store the verification tables for this purpose. Later in 1998, Jan and Chen proposed a password-based authentication scheme [12]. However, their scheme does not require to store the verification tables in the system as compared with Lamport's scheme. Jan and Chen's scheme is ineffective in the sense that the server needs to maintain the verification tables and thus the size of the verification tables are proportional to the number of users. In 2000, Hwang and Li proposed a remote user authentication scheme with the help of smart cards [13]. Their scheme is based on ElGamal public-key cryptosystem [14]. Hwang and Li's scheme does not require to store password table in the system and it only needs to maintain one secret key. As a result, Hwang and Li's scheme is effective compared with [4], [12].

In 2002, Lee et al. proposed a fingerprint-based remote user authentication scheme using smart cards [15]. However, their scheme is vulnerable to impersonation attack [8], [16]. In 2004, Lin and Lai proposed a flexible biometric-based remote user authentication scheme [8]. Later in 2007, Khan and Zhang showed that Lin and Lai's scheme is insecure against the server spoofing attack [17]. In 2010, Li-Hwang proposed an efficient biometric-based remote user authentication scheme using smart cards [6]. Their scheme is based on biometrics verification, smart card and one-way hash function, and it also uses the random nonce rather than a synchronized clock, and thus it is very efficient in computational cost. However, recently in 2011, Li et al. showed that Li-Hwang's scheme is insecure because Li-Hwang's scheme fails to provide proper authentication and also it is vulnerable to man-in-the-middle attacks [7]. Unfortunately, in this paper we show that Li et al.'s scheme [7] again fails to provide proper authentication in login and authentication phases because there is no verification on user's entered password after successful verification of his/her biometric template. We also show that due to the same password verification problem Li et al.'s scheme fails to update the new password correctly of a user in password change phase.

The organization of the paper is as follows. In Section 2, we briefly review Li et al.'s biometric-based remote user authentication scheme using smart cards [7]. We then show the security weaknesses in their scheme in Section 3. In Section 4, we propose an improvement of their scheme in order to eliminate weaknesses discussed in Section 3 and then compare performance and functionality of our improved scheme with Li et al.'s scheme. Finally, we conclude the paper in Section 5.

International Journal of Network Security & Its Applications (IJNSA), Vol.3, No.2, March 2011

## 2. REVIEW OF LI ET AL.'S BIOMETRIC-BASED REMOTE USER AUTHENTICATION SCHEME

In this section, for a detailed analysis we review in brief Li et al.'s efficient biometric-based remote user authentication scheme [7]. The notations used throughout this paper are summarized in Table 1.

Table 1. Notations used in this paper.

| Symbol | Description |
| --- | --- |
| $C_i$ | Client (user) |
| $R$ | Trusted registration center |
| $S_i$ | Server |
| $PW_i$ | $C_i$'s password |
| $ID_i$ | Identity of $C_i$ |
| $SID_i$ | Identity of $S_i$ |
| $B_i$ | Biometric template of $C_i$ |
| $h(\cdot)$ | A secure one-way hash function |
| $X_s$ | The secret information (the master secret key) maintained by the server |
| $R_c$ | A secret number chosen by $C_i$ used to prevent replay attacks |
| $R_s$ | A secret number chosen by $S_i$ used to prevent replay attacks |
| $y$ | A secret number maintained by $S_i$ and stored in user's smart card |
| $\|$ | Message concatenation operation |
| $\oplus$ | Exclusive-or operation |

A one-way hash function $h:\{0,1\}^* \rightarrow \{0,1\}^l$ takes an arbitrary-length input $X \in \{0,1\}^*$, and produces a fixed-length (say, $l$-bits) output $h(X) \in \{0,1\}^l$, called the message digest. The hash function is the fingerprint of a file, a message, or other data blocks, and has the following attributes [11].

- $X$ can be applied to a data block of all sizes.
- For any given variable $X$, $h(X)$ is easy to operate, enabling easy implementation in software and hardware.
- The output length of $h(X)$ is fixed.
- Deriving $X$ from the given value $Y = h(X)$ and the given hash function $h(\cdot)$ is computationally infeasible.
- For any given variable $X$, finding any $Y \neq X$ so that $h(Y) = h(X)$ is computationally infeasible.
- Finding a pair of inputs $(X, Y)$, with $X \neq Y$, so that $h(X) = h(Y)$ is computationally infeasible.

The hash functions have many applications in the field of cryptology and information security, notably in digital signatures, message authentication codes (MACs), and other form of authentications and thus it becomes the basis of many cryptographic protocols. One of the fundamental properties of hash functions is that the outputs are very sensitive to small perturbations in their inputs and in general, those cryptographic hash functions cannot be applied straightforwardly when the input data with noisy such as biometrics [9]. If there are few differences between the input each time and this situation will cause the legal user unable to pass biometric authentication. As in [7], we also use the biometric template matching to perform the biometric authentication in our improved scheme.

Li et al.'s scheme consists of the following four phases, namely, *registration phase*, *login phase*, *authentication phase* and *password change phase*.

### 2.1. Registration Phase

Before the remote user $C_i$ logins to the system, $C_i$ needs to perform the following steps:
- Step 1: $C_i$ chooses a random number $N$ and computes the masked password $RPW_i = h(N \parallel PW_i)$ based on his/her password $PW_i$. $C_i$ then enters his/her personal biometrics $B_i$ on the specific device and provides his/her masked password $RPW_i$, the identity $ID_i$ of $C_i$ to the registration center, $R$ via a secure channel.
- Step 2: After receiving the information in Step 1, $R$ computes $r_i$ and $e_i$ as $r_i = h(RPW_i \parallel f_i)$ and $e_i = h(ID_i \parallel X_s) \oplus r_i$, where $f_i = h(B_i)$.
- Step 3: $R$ stores $(f_i, e_i, h(\cdot), y)$ on $C_i$'s smart card and sends it to $C_i$ via secure channel.
- Step 4: $C_i$ also enters $N$ into his/her smart card.

### 2.2. Login Phase

When the user $C_i$ wants to login to the remote server $S_i$, he/she needs to perform the following steps:
- Step 1: User $C_i$ first inserts his/her smart card into the card reader and provides his/her personal biometrics $B_i$ on the specific device to verify the user's biometrics. If the biometrics information matches the template stored in the system, $C_i$ passes the biometric verification and then performs the following steps.
- Step 2: $C_i$ now inputs his/her password $PW_i$ and identity $ID_i$. After receiving $C_i$'s $ID_i$ and $PW_i$ the smart card computes $RPW_i = h(N \parallel PW_i), r'_i = h(RPW_i \parallel f_i)$, $M_1 = e_i \oplus r'_i, M_2 = M_1 \oplus R_c$, where $R_c$ is a random number generated by the user, $M_3 = h(y \parallel R_c), M_4 = RPW_i \oplus M_3, M_5 = h(M_2 \parallel M_3 \parallel M_4)$.
- Step 3: Finally, $C_i$ sends the message $(ID_i, M_2, M_4, M_5)$ to server $S_i$.

### 2.3. Authentication Phase

Once the request login message $(ID_i, M_2, M_4, M_5)$ from the user $C_i$ is received, the remote server $S_i$ and the user $C_i$ perform the following steps for mutual authentication.
- Step 1: $S_i$ first checks the format of $ID_i$. If it is valid, $S_i$ then computes $M_6 = h(ID_i \parallel X_s), M_7 = M_2 \oplus M_6, M_8 = h(y \parallel M_7)$ and then verifies whether $M_5 = h(M_2 \parallel M_8 \parallel M_4)$. If it holds, $S_i$ stores $(ID_i, M_7)$ in its database. Thus, when $S_i$ received the next $C_i$'s login message, $S_i$ computes $M'_7$ and compares with the stored $M_7$ in the database. If these values are same, $S_i$ rejects it because it is a replay message. Otherwise, $S_i$ replaces the old $M_7$ by the new computed $M_7'$. This method resists the replay attacks and man-in-the-middle attacks compared to Li-Hwang's scheme [6].

- Step 2: If Step 1 does not hold, $S_i$ rejects the login request and terminates the session. Otherwise, $S_i$ accepts the login request and thus $C_i$ is authenticated by $S_i$ as a valid user. $S_i$ then computes $M_9 = M_4 \oplus M_8, M_{10} = h(M_9 \parallel SID_i \parallel y) \oplus M_8 \oplus R_s$, $M_{11} = h(M_6 \parallel M_9 \parallel y \parallel R_s)$, where $R_s$ is a random number chosen by $S_i$. $S_i$ sends the message $(M_{10}, M_{11})$ to $C_i$.

- Step 3: Upon receiving the message in Step 2, $C_i$ then computes $M_{12} = h(RPW_i \parallel SID_i \parallel y) \oplus M_3 \oplus M_{10}$ and goes for verification of the equality $M_{11} = h(M_1 \parallel RPW_i \parallel y \parallel M_{12})$. If this verification holds good, the validity of $S_i$ is authenticated by $C_i$. Otherwise, $C_i$ terminates the scheme.

- Step 4: After mutual authentication phase, the user $C_i$ and the server $S_i$ compute their session key $SK$ as $SK = h(RPW_i \parallel M_3 \parallel M_{12} \parallel SID_i)$ (by $C_i$) and $SK = h(M_9 \parallel M_8 \parallel R_s \parallel SID_i)$ (by $S_i$).

## 2.4. Password Change Phase

To freely change the existing password $PW_i$ of a user $C_i$ by a new password $PW_i^{new}$, the user $C_i$ does the following steps.

$C_i$ first inserts his/her smart card into the card reader and provides his/her personal biometrics $B_i$ on the specific device to verify the user's biometrics. If this verification passes, the user $C_i$ enters his/her old password $PW_i$ and new password $PW_i^{new}$. After that, the smart card performs the following operations: $RPW_i' = h(N \parallel PW_i)$, $r_i' = h(RPW_i' \parallel f_i)$, $e_i' = e_i \oplus r_i'$, where $f_i = h(B_i)$, $RPW_i^{new} = h(N \parallel PW_i^{new})$, $r_i'' = h(RPW_i^{new} \parallel f_i)$ and $e_i^{new} = e_i' \oplus r_i''$. After these operations, the value of $e_i$ is replaced by the new calculated value of $e_i^{new}$ in the smart card.

## 3. CRYPTANALYSIS OF LI ET AL.'S BIOMETRIC-BASED REMOTE USER AUTHENTICATION SCHEME

In this section, we show that the Li et al.'s scheme [7] has the following security weaknesses in their design.

### 3.1. Li et al.'s Scheme fails to provide strong authentication in Login and Authentication Phases

From login phase of Li et al.'s scheme, it is noted that user $C_i$ first enters his/her personal biometrics on the specific device to verify whether his/her biometrics passes or not. If this verification passes, then $C_i$ enters his/her password $PW_i$ and identity $ID_i$. In both Li-Hwang's scheme [6] and Li et al.'s scheme [7], it is assumed that the user $C_i$ always enters his/her

17

password $PW_i$ correctly. There is no verification on entered password $PW_i$ in login phase. In practice, a user keeps different passwords for different purposes. Even if the user $C_i$ enters his/her password incorrectly by mistake, both the login and authentication phases still continue. Finally, at the end of authentication phase, $S_i$ rejects $C_i$'s login request and as a result, we have unnecessarily extra communication and computational overheads during login and authentication phases. In such scenario, the user $C_i$ is totally unaware of the fact that he/she has entered his/her password incorrectly in login phase.

In our cryptanalysis, we assume that the user $C_i$ enters his/her password incorrectly by mistake and let the entered password be $PW_i'(\neq PW_i)$. Then from the login phase, it follows that

$$RPW_i' = h(N \| PW_i')$$
$$\neq h(N \| PW_i),$$
$$r_i' = h(RPW_i' \| f_i)$$
$$= h(h(N \| PW_i') \| f_i)$$
$$\neq h(h(N \| PW_i) \| f_i).$$

As a result, we have the following computations:

$$M_1 = e_i \oplus r_i'$$
$$= h(ID_i \| X_s) \oplus h(h(N \| PW_i) \| f_i) \oplus h(h(N \| PW_i') \| f_i)$$
$$\neq h(ID_i \| X_s),$$

$$M_2 = M_1 \oplus R_c$$
$$= h(ID_i \| X_s) \oplus h(h(N \| PW_i) \| f_i) \oplus h(h(N \| PW_i') \| f_i) \oplus R_c$$
$$\neq h(ID_i \| X_s) \oplus R_c,$$

$$M_3 = h(y \| R_c),$$
$$M_4 = RPW_i' \oplus M_3$$
$$= h(N \| PW_i') \oplus h(y \| R_c)$$
$$\neq h(N \| PW_i) \oplus h(y \| R_c),$$
$$M_5 = h(M_2 \| M_3 \| M_4)$$
$$\neq h((h(ID_i \| X_s) \oplus R_c) \| h(y \| R_c) \| (h(N \| PW_i) \oplus h(y \| R_c))).$$

After sending the message $(ID_i, M_2, M_4, M_5)$ to $S_i$ by $C_i$, the server $S_i$ will compute $M_7$ and $M_8$ incorrectly in authentication phase after verifying the format of $ID_i$, because

$$M_6 = h(ID_i \oplus X_s),$$
$$M_7 = M_2 \oplus M_6$$
$$\neq h(ID_i \| X_s) \oplus R_c \oplus h(ID_i \| X_s)$$
$$\neq R_c.$$

Thus, we have,

$$M_8 = h(y \parallel M_7),$$
$$\neq h(y \parallel R_c),$$

and hence, when $S_i$ compares $M_5$ with $h(M_2 \parallel M_8 \parallel M_4)$, it is obvious that

$$M_5 \neq h(M_2 \parallel M_8 \parallel M_4).$$

$S_i$ does not store $(ID_i, M_7)$ in the database and it rejects the login request and terminates the session. Thus, $C_i$ is not authenticated as a valid user by $S_i$. Here the user $C_i$ is not notified that he/she has entered his/her password incorrectly in login phase.

## 3.2. Li et al.'s Scheme fails to update properly User's new password locally in Password Change Phase

In Li et al.'s biometric-based remote user authentication scheme, any user $C_i$ can freely change his/her old password by the new password locally without the help of the registration center $R$. However, in their scheme only the verification of biometrics of user $C_i$ takes place and in case the verification passes, $C_i$ is allowed to input his/her old password $PW_i^{old}$ and new password $PW_i^{new}$. Since there is no verification of old password in their scheme, the updation of new password will take place incorrectly if the user $C_i$ enters his/her old password $PW_i^{old}$ wrongly.

In order to update the new password, Li et al.'s scheme does the following steps after successful verification of user's biometric template $B_i$. We assume that the user $C_i$ enters his/her old password $PW_i^{old}$ incorrectly so that $PW_i \neq PW_i^{old}$.

We then have,

$$RPW_i' = h(N \parallel PW_i^{old})$$
$$\neq h(N \parallel PW_i),$$
$$r_i' = h(RPW_i' \parallel f_i)$$
$$= h(h(N \parallel PW_i^{old}) \parallel f_i)$$
$$\neq (h(N \parallel PW_i) \parallel f_i), and$$
$$e_i' = e_i \oplus r_i'$$
$$= h(ID_i \parallel X_s) \oplus r_i \oplus r_i'$$
$$= h(ID_i \parallel X_s) \oplus h(h(N \parallel PW_i) \parallel f_i) \oplus h(h(N \parallel PW_i^{old}) \parallel f_i)$$
$$\neq h(ID_i \parallel X_s).$$

As a result, we have the following result,

$$RPW_i^{new} = h(N \parallel PW_i^{new}),$$
$$r_i'' = h(RPW_i^{new} \parallel f_i),$$
$$e_i^{new} = e_i' \oplus r_i''$$
$$\neq h(ID_i \parallel X_s) \oplus h(RPW_i^{new} \parallel f_i).$$

The smart card replaces $e_i$ with $e_i^{new}$ in its memory. Obviously, in such scenario the new password of user $C_i$ will not be updated correctly in the smart card. We see that due to this problem, when the user logins later in the system providing his/her biometrics as well as new password $PW_i^{new}$, the login request of the user is always rejected by $S_i$ even if in that time the user enters new password correctly. This effect also continues in subsequent password change phases by the user $C_i$. Thus, to overcome from this serious problem the only solution is to issue a new smart card with necessary information from the user $C_i$'s fresh identity and masked password provided to the registration server $R$.

## 4. THE PROPOSED IMPROVED SCHEME

In this section, we propose an improvement based on Li et al.'s biometric-based remote user authentication scheme using smart cards. The improved scheme keeps the merit of the original Li et al.'s scheme and can withstand the original scheme's security weaknesses discussed in Section 3.

### 4.1. Description of the Improved Scheme

We discuss the registration, login, authentication and password change phases of our proposed scheme.

#### 4.1.1. Registration Phase

When the remote user authentication scheme starts, the user $C_i$ and the registration center $R$ need to perform the following steps:
- Step 1: $C_i$ chooses a random number $N$ and computes the masked password $RPW_i = h(N \parallel PW_i)$ based on his/her password $PW_i$. $C_i$ enters his/her personal biometrics $B_i$ on the specific device and provides his/her masked password $RPW_i$, the identity $ID_i$ of $C_i$ to the registration center, $R$ via a secure channel.
- Step 2: After receiving the above information in Step 1, the registration center $R$ computes $r_i$ and $e_i$ as $r_i = h(RPW_i \parallel f_i)$ and $e_i = h(ID_i \parallel X_s) \oplus r_i$, where $f_i = h(B_i)$.
- Step 3: $R$ stores $(f_i, r_i, e_i, h(\cdot), y)$ on $C_i$'s smart card and sends it to $C_i$ via secure channel.
- Step 4: Finally, $C_i$ enters $N$ into his/her smart card.

It is noted that in our improved scheme, the extra information $r_i$ is stored in the user's smart card compared with Li et al.'s scheme. The registration phase is summarized in Figure 1.

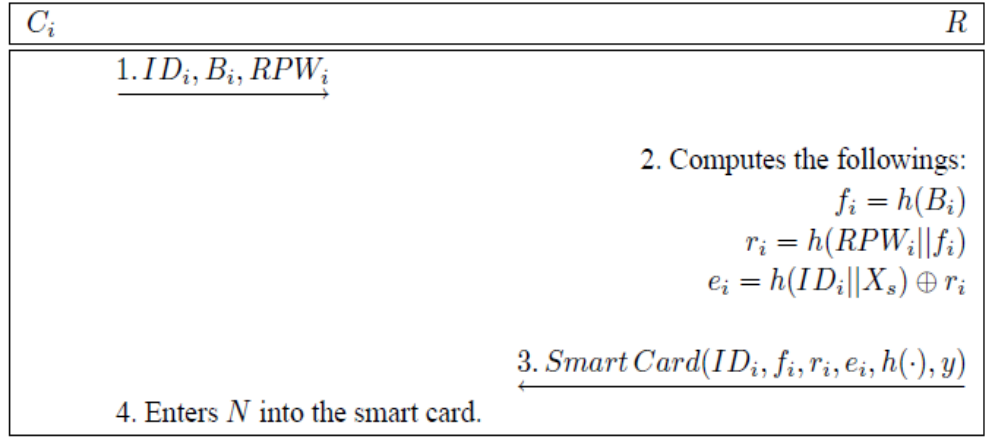

**Fig. 1.** Registration phase of the proposed scheme.

### 4.1.2. Login Phase

When the user $C_i$ wants to login to the remote server $S_i$, he/she needs to perform the following steps:

- Step 1: User $C_i$ first inserts his/her smart card into the card reader and provides his/her personal biometrics $B_i$ on the specific device to verify the user's biometrics $B_i$. If the biometrics information matches the template stored in the system, $C_i$ passes the biometric verification and then performs the following steps.
- Step 2: $C_i$ inputs his/her password $PW_i$ and identity $ID_i$. After receiving $C_i$'s $ID_i$ and $PW_i$ the smart card computes $RPW_i = h(N \parallel PW_i), r'_i = h(RPW_i \parallel f_i)$.
- Step 3: The smart card then checks whether $r_i = r'_i$. If this verification does not hold, this means that the user $C_i$ enters his/her password incorrectly. The user is notified with the incorrect password error message and the scheme terminates.
- Step 4: Otherwise, if $r_i = r'_i$, then only the smart card computes $M_1 = e_i \oplus r'_i, M_2 = M_1 \oplus R_c$, where $R_c$ is a random number generated by the user, $M_3 = h(y \parallel R_c), M_4 = RPW_i \oplus M_3, M_5 = h(M_2 \parallel M_3 \parallel M_4)$.
- Step 5: Finally, $C_i$ sends the message $\langle ID_i, M_2, M_3, M_4, M_5 \rangle$ to server $S_i$.

The login phase is summarized in Figure 2.

### 4.1.3. Authentication Phase

After receiving the request login message $\langle ID_i, M_2, M_3, M_4, M_5 \rangle$ from the user $C_i$, the remote server $S_i$ and the user $C_i$ perform the following steps for mutual authentication.

- Step 1: $S_i$ first checks the format of $ID_i$. If it is valid, $S_i$ then computes $M_6 = h(ID_i \parallel X_s), M_7 = M_2 \oplus M_6, M_8 = h(y \parallel M_7)$ and then verifies whether $M_8 = M_3$. If it holds, then $S_i$ further verifies whether $M_5 = h(M_2 \parallel M_8 \parallel M_4)$. If it holds, $S_i$ stores $(ID_i, M_7)$ in its database. Thus, when $S_i$ received the next $C_i$'s login message, $S_i$ computes $M'_7$ and compares with the stored $M_7$ in the database. If these values are same, $S_i$ rejects it because it is a replay message. Otherwise, $S_i$ replaces the old $M_7$ by the new computed $M'_7$. This method resists the replay attacks and man-in-the-middle attacks compared to Li-Hwang's scheme [6].

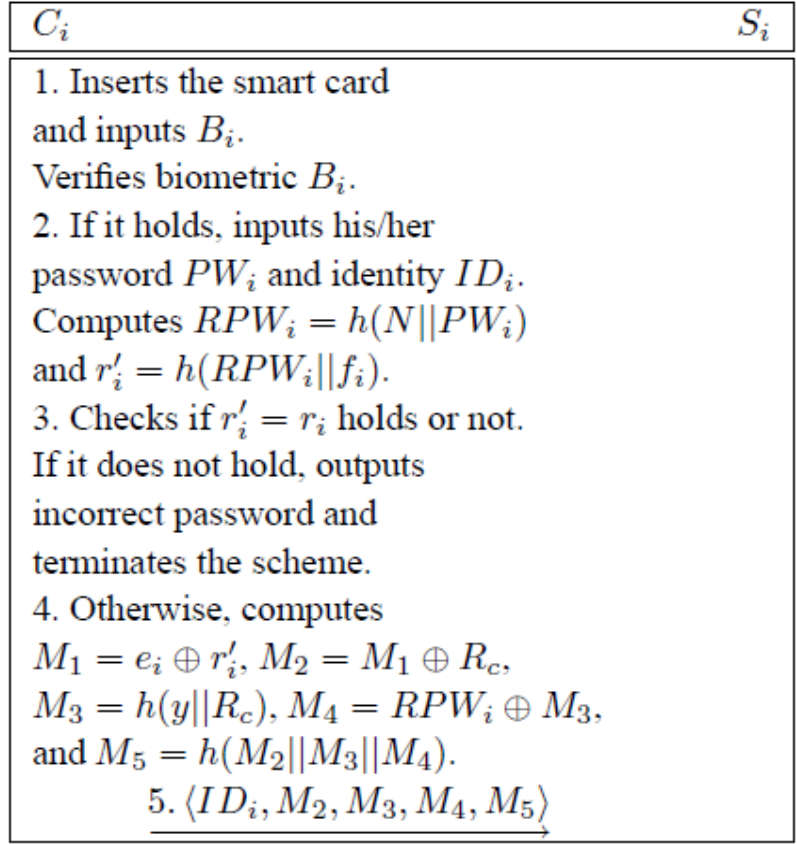

Fig. 2. Login phase of the proposed scheme.

- Step 2: If Step 1 does not hold, this means that $C_i$ is not legitimate and hence, $S_i$ rejects the login request and terminates the session. Otherwise, $S_i$ accepts the login request and thus $C_i$ is authenticated by $S_i$ as a valid user. $S_i$ then computes $M_9 = M_4 \oplus M_8$, $M_{10} = h(M_9 \parallel SID_i \parallel y) \oplus M_8 \oplus R_s$, $M_{11} = h(y \parallel R_s)$, $M_{12} = h(M_6 \parallel M_9 \parallel y \parallel R_s)$, where $R_s$ is a random number chosen by $S_i$. $S_i$ sends the message $\langle M_{10}, M_{11}, M_{12} \rangle$ to $C_i$.
- Step 3: After receiving the message in Step 2, $C_i$ then computes $M_{13} = h(RPW_i \parallel SID_i \parallel y) \oplus M_3 \oplus M_{10}$. $C_i$ computes $M_{14} = h(y \parallel M_{13})$, and

22

verifies whether $M_{14} = M_{11}$. If it holds, then $C_i$ further goes for verification of the equality $M_{12} = h(M_1 \parallel RPW_i \parallel y \parallel M_{13})$. If this verification holds good, the validity of $S_i$ is authenticated by $C_i$. Otherwise, $C_i$ terminates the scheme.

- Step 4: After mutual authentication phase, the user $C_i$ computes the session key $SK$ shared with the server $S_i$ as $SK = h(RPW_i \parallel M_3 \parallel M_{13} \parallel SID_i)$. The server $S_i$ also computes the same session key $SK$ shared with the user $C_i$ as $SK = h(M_9 \parallel M_8 \parallel R_s \parallel SID_i)$.

The authentication phase is summarized in Figure 3.

| $C_i$ | $S_i$ |
|---|---|
|  | 1. Checks the format of $C_i$'s $ID_i$ |
|  | 2. Computes $M_6 = h(ID_i\|\|X_s)$, $M_7 = M_2 \oplus M_6$, $M_8 = h(y\|\|M_7)$ and verifies whether $M_8 = M_3$. |
|  | 3. If it holds, further verifies whether $M_5 = h(M_2\|\|M_8\|\|M_4)$. If it holds, $S_i$ stores $(ID_i, M_7)$ in its database. $S_i$ computes $M_9 = M_4 \oplus M_8$, $M_{10} = h(M_9\|\|SID_i\|\|y) \oplus M_8 \oplus R_s$, $M_{11} = h(y\|\|R_s)$, $M_{12} = h(M_6\|\|M_9\|\|y\|\|R_s)$ |
|  | 4. $\langle M_{10}, M_{11}, M_{12}\rangle$ ← |
| 5. Computes $M_{13} = h(RPW_i\|\|SID_i\|\|y) \oplus M_3 \oplus M_{10}$, $M_{14} = h(y\|\|M_{13})$ and verifies if $M_{14} = M_{11}$. It it holds, verifies $M_{12} = h(M_1\|\|RPW_i\|\|y\|\|M_{13})$ 6. It above holds, $S_i$ is authenticated by $C_i$. 7. Computes $SK = h(RPW_i\|\|M_3\|\|M_{13}\|\|SID_i)$ |  |
|  | 8. Computes $SK = h(M_9\|\|M_8\|\|R_s\|\|SID_i)$ |

**Fig. 3.** Authentication phase of the proposed scheme.

### 4.1.4. Password Change Phase

In Li et al.'s scheme a user can freely change his/her password locally without contacting the registration center. In their scheme, only after the successful biometric template verification, even if the user enters by mistake his/her old password wrongly the value of $e_i$ is still updated,

23

which is incorrect. Thus, in next time when the user wants to login in the system, the user's login request will be rejected by the server even the user enters his/her new changed password correctly in that time.

In this phase, we show that in our improved scheme the above mentioned situation will never happen because the smart card always verifies the old entered password by the user before updating the new changed password. The password change phase consists of the following steps:

- Step 1: $C_i$ enters the smart card into the card reader and offers his/her personal biometrics $B_i$ on the specific device. If the biometric information matches the template stored in system, $C_i$ passes the biometric verification, and $C_i$ performs the following steps.

- Step 2: $C_i$ enters his/her old password $PW_i^{old}$ and new changed password $PW_i^{new}$.

- Step 3: The smart card then computes the following:

  $RPW_i' = h(N \parallel PW_i^{old}), r_i' = h(RPW_i' \parallel f_i).$

  If $r_i' \neq r_i$, it means that $C_i$ enters his/her old password incorrectly and the password change phase is terminated.

- Step 4: On the other hand, if $r_i' = r_i$, then only the smart card computes

  $$e_i' = e_i \oplus r_i'$$
  $$= h(ID_i \parallel X_s) \oplus r_i \oplus r_i'$$
  $$= h(ID_i \parallel X_s),$$
  $$RPW_i^{new} = h(N \parallel PW_i^{new}),$$
  $$r_i'' = h(RPW_i^{new} \parallel f_i),$$
  $$e_i'' = e_i' \oplus r_i''$$
  $$= h(ID_i \parallel X_s) \oplus h(RPW_i^{new} \parallel f_i).$$

- Step 5: Finally, replace $e_i$ with $e_i''$ and $r_i$ with $r_i''$ in the smart card.

We note that in the proposed scheme not only the $e_i$ is replaced with $e_i''$ in the smart card, but also the $r_i$ is replaced with $r_i''$ in the smart card as compared with Li et al.'s scheme.

## 4.2. Analysis of the Improved Scheme

In this section, we discuss the security features of the proposed scheme and then compare the performances of the proposed scheme with those for Li et al.'s scheme [7].

### 4.2.1. Security Analysis

The proposed scheme resists the man-in-the-middle attacks as in Li et al.'s scheme, because the proposed scheme also retains the merit of the original Li et al.'s scheme. We discuss in the

International Journal of Network Security & Its Applications (IJNSA), Vol.3, No.2, March 2011

following the security features of our proposed scheme with respect to stolen smart card attacks, strong mutual authentication, replay attacks, security of session key and security of secret key.

- **Resist stolen smart card attacks** Suppose the user's smart card has been lost or stolen and the attacker can breach the information $(ID_i, f_i, r_i, e_i, h(\cdot), y, N)$ which are stored in the smart card. Now knowing $r_i = h(h(N \| PW_i) \| f_i)$, where $f_i = h(B_i)$ and $e_i = h(ID_i \| X_s) \oplus r_i$, the attacker can compute $h(ID_i \| X_s) = e_i \oplus r_i$. However, due to properties of one-way hash function $h(\cdot)$ it is computationally infeasible to compute the master secret key $X_s$ from the exacted $h(ID_i \| X_s)$. Again, the attacker cannot get the user's password $PW_i$ from $r_i = h(h(N \| PW_i) \| f_i)$ due to properties of one-way hash function $h(\cdot)$. Besides, it is difficult for the attacker to pass the biometrics verification and so the attacker cannot derive or change the user's password $PW_i$. Moreover, the password $PW_i$ is unknown to the registration center $R$. As a result, the proposed scheme can resist insider attacks.

- **Strong mutual authentication** In the login phase, while constructing $M_1 = e_i \oplus r_i'$ the smart card always verifies the user's entered password. If the password verification passes, $C_i$ sends the message $\langle ID_i, M_2, M_3, M_4, M_5 \rangle$ to server $S_i$. $S_i$ then checks the validity of user's ID, $ID_i$, and computes $M_6 = h(ID_i \| X_s), M_7 = M_2 \oplus M_6$, $M_8 = h(y \| M_7) = h(y \| R_c)$. $S_i$ can authenticate $C_i$ by verifying equations $M_8 = M_3$ and $M_5 = h(M_2 \| M_8 \| M_4)$. If these hold good (since the user's password is authenticated in the login phase), $C_i$ is a valid user. Since the authentication relies on the one-way hash function $h(\cdot)$, any fabricated message $\langle ID_i, M_2', M_3', M_4', M_5' \rangle$ cannot pass the authentication procedure in the proposed scheme. Similarly, when $S_i$ sends the message $\langle M_{10}, M_{11}, M_{12} \rangle$ to $C_i$, $C_i$ then computes $M_{13} = h(RPW_i \| SID_i \| y) \oplus M_3 \oplus M_{10}$. $C_i$ also computes $M_{14} = h(y \| M_{13})$, and verifies whether $M_{14} = M_{11}$ holds or not. If it holds, then $C_i$ further goes for verification of the equality $M_{12} = h(M_1 \| RPW_i \| y \| M_{13})$. If this verification holds good, the validity of $S_i$ is authenticated by $C_i$. Otherwise, $C_i$ terminates the scheme. Also, for the same reason any fabricated message $\langle M_{10}', M_{11}', M_{12}' \rangle$ cannot pass the authentication procedure in the proposed scheme. So, the proposed scheme provides strong mutual authentication.

- **Resist replay attacks** In our scheme, an attacker may try to attempt to pretend to be a valid user to login the server by sending messages previously transmitted by a legal user. However, the proposed scheme uses random nonces and it also compares $M_7$ value in its database to withstand the replay attacks. Since the random nonces $R_c$ and $R_s$ are generated by the user and server independently, and both values are distinct due to random selection in each session, an attacker has no way to successfully replay used messages.

25

- *Security of secret key* In our proposed scheme, as in [7] there are two pieces of secret information used: the master secret key $X_s$ is shared between $R$ and $S_i$, and the secret number $y$ which is maintained by $S_i$ and stored in the user's smart card. In order to derive $X_s$ and $y$ from the transmitted messages $\langle ID_i, M_2, M_3, M_4, M_5 \rangle$ and $\langle M_{10}, M_{11}, M_{12} \rangle$, it is computationally infeasible problem due to one-way properties of the hash function $h(\cdot)$. Hence, there is no way for an attacker to compute the secret session key $SK$ shared between $C_i$ and $S_i$.

- *Security of session key*

  - *Known-key secrecy* It means that if an attacker is able to compromise one session key, that does not lead to compromise other session keys. Assume that the attacker has the session key $SK = h(RPW_i \parallel M_3 \parallel M_{13} \parallel SID_i) = h(M_9 \parallel M_8 \parallel R_s \parallel SID_i)$. Now this session key is associated with the random nonces $R_c$, $R_s$, the masked password of user $RPW_i = h(N \parallel PW_i)$ and the secret number $y$. It is noted that all these values are protected by the one-way hash function $h(\cdot)$. Hence, even if the attacker compromises a past session key, he/she cannot compute the random nonces $R_c$, $R_s$, the masked password of user $RPW_i = h(N \parallel PW_i)$ and the secret number $y$ from the compromised session key and thus, he/she cannot derive other session keys.

  - *Forward secrecy* This means that if somehow the master key of the system is compromised, the secrecy of the previously established session keys should not be affected. Now, if the master secret key $X_s$ is compromised for some reasons, the attacker cannot compute any previously established secret keys without having $RPW_i = h(N \parallel PW_i)$, $y$ and $R_s$. Hence, as in [7] the proposed scheme also provides forward secrecy.

### 4.2.2. Functionality Analysis

Functionality comparison of our improved scheme with Li et al.'s scheme is shown in Table 2. Due to computational efficiency of hash function, both our scheme and Li et al.'s scheme are very efficient in computation. The proposed scheme provides strong authentication by verifying user's personal biometrics, password, and random nonces generated by the user and server as compared to that for Li et al.'s scheme. Besides this advantage, the proposed scheme supports changing password by the user locally without the help of registration center and updates always correctly the new password replacing the old password in the smart card as compared to that for Li et al.'s scheme. The proposed scheme only requires additional storage requirement due to storing $r_i$ in the memory of the smart card compared with Li et al.'s scheme. Due to computational efficiency of the hash function $h(\cdot)$ our scheme is also efficient as in Li et al.'s scheme. Moreover, the proposed scheme supports other functionality as in Li et al.'s scheme.

Table 2. Functionality comparison between our proposed scheme and Li et al.'s scheme

| | Our scheme | Li et al.'s scheme |
|---|---|---|
| Storage cost | $ID_i + f_i + e_i + h(\cdot) + y + N + r_i$ | $ID_i + f_i + e_i + h(\cdot) + y + N$ |
| Computation cost | $17T_h$ | $15T_h$ |
| Strong authentication | Yes | No |
| Session key agreement | Yes | Yes |
| Correct password change freely | Yes | No |
| No time synchronization | Yes | Yes |
| Resist stolen smart card attacks | Yes | Yes |
| Resist man-in-the-middle attacks | Yes | Yes |
| Resist replay attacks | Yes | Yes |
| Insecure communication | Two times | Two times |

Note: $T_h$ denotes one-way hashing operation.

# 5. CONCLUSIONS

In this paper, we have shown that Li et al.'s scheme fails to provide strong authentication in both login and authentication phases. We have further shown that Li et al.'s scheme fails to update properly the user's changed password locally without the help of the registration center. We have then proposed an improvement of Li et al.'s biometric-based remote user authentication scheme using smart cards in order to withstand the above security weaknesses in their scheme. The proposed scheme also keeps the merit of the original Li et al.'s scheme. Our proposed scheme provides strong authentication with the help of verifying biometrics, passwords and random nonces generated by the user and server as compared to that for Li et al.'s scheme. Further, the proposed scheme always updates the new password correctly in the smart card changed by the user freely without the help of registration center.

## ACKNOWLEDGEMENTS

The author would like to thank the anonymous reviewers for their constructive suggestions and comments which have improved the content as well as presentation of this paper.

**Author**

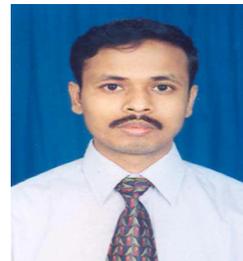

**Dr. Ashok Kumar Das** is currently working as an Assistant Professor in the Center for Security, Theory and Algorithmic Research of the International Institute of Information Technology (IIIT), Hyderabad 500 032, India. Prior to joining IIIT Hyderabad, he held academic position as an Assistant Professor in Department of Computer Science and Engineering of the International Institute of Information Technology, Bhubaneswar 751 013, India from July 2008 to May 2010. He received his Ph.D. degree in Computer Science and Engineering from the Indian Institute of Technology, Kharagpur, India on April 2009. He received the M.Tech. degree in Computer Science and Data Processing from the Indian Institute of Technology, Kharagpur, India on January 2000. He also received the M.Sc. degree in Mathematics from the Indian Institute of Technology, Kharagpur, India, in 1998. Prior to join in Ph.D., he worked with C-DoT (Centre for Development of Telematics), a premier telecom technology centre of Govt. of India at New Delhi, India from March 2000 to January 2004. His current research interests include cryptography, security in wireless sensor networks, mobile adhoc networks and vehicular adhoc networks, proxy ring signature and remote user authentication. He has published over 20 papers in international journals and conferences in these areas.